\documentclass[prb,twocolumn,superscriptaddress]{revtex4}

\pdfoutput=1

\usepackage{graphicx}
\usepackage{amsmath}
\usepackage{amsfonts}
\usepackage{color}
\usepackage{bm}

\begin{document}

\title{A quantum dot single-photon source with on-the-fly all-optical \\ polarization control and timed emission}

\author{Dirk Heinze}

\affiliation{Physics Department and Center for Optoelectronics and Photonics Paderborn (CeOPP), Universit\"at Paderborn, Warburger Strasse 100, 33098 Paderborn, Germany}

\author{Artur Zrenner}

\affiliation{Physics Department and Center for Optoelectronics and Photonics Paderborn (CeOPP), Universit\"at Paderborn, Warburger Strasse 100, 33098 Paderborn, Germany}

\author{Stefan Schumacher}

\affiliation{Physics Department and Center for Optoelectronics and Photonics Paderborn (CeOPP), Universit\"at Paderborn, Warburger Strasse 100, 33098 Paderborn, Germany}

\affiliation{College of Optical Sciences, University of Arizona, Tucson, Arizona 85721, USA}

\date{\today}




\maketitle

\textbf{Sources of single photons are key elements in the study of basic quantum optical concepts and applications in quantum information science. Among the different sources available, semiconductor quantum dots excel with their straight forward integrability in semiconductor based on-chip solutions\cite{Stock2012,Reithmaier2013} and the potential that photon emission can be triggered on demand.\cite{Michler2000} Usually, the photon emission event is part of a cascaded biexciton-exciton emission scheme. Important properties of the emitted photon such as polarization and time of emission are either probabilistic in nature or pre-determined by electronic properties of the system.\cite{strauf2007} In this work, we study the direct two-photon emission from the biexciton.\cite{Ota2011} We show that emission through this higher-order transition provides a much more versatile approach to generate a single photon. In the scheme we propose, the two-photon emission from the biexciton is enabled by a laser field (or laser pulse) driving the system into a virtual state inside the band gap. From this intermediate virtual state, the single photon of interest is then spontaneously emitted. Its properties are determined by the driving laser pulse, enabling all-optical on-the-fly control of polarization state, frequency, and time of emission of the photon.}

Semiconductor quantum dots have proven their promise as basic building blocks in various applications in the field of semiconductor based quantum optics and quantum communication.\cite{bookJahnke2012} These semiconductor nanostructures have been used as well-controlled on demand quantum emitters for single photons \cite{Michler2000,strauf2007,Mehta2010,SoniaBuckley2012,Prechtel2013,Matthiesen2013,He2013,Wei2014} as well as for lasing at the single-photon level\cite{Wiersig2009,Strauf2011}  and to generate polarization entangled pairs of photons.\cite{Benson2000,Stevenson2006b,Akopian2006,Hafenbrak2007,Dousse2010,Mueller2014} 
It has been demonstrated that the interaction of the quantum dot's electronic excitations with optical fields and the emission characteristics of photons can be tailored to a large extend by use of optical cavities.\cite{SoniaBuckley2012} Even on-chip solutions of quantum-dot cavity systems with build-in electrically pumped microlaser sources have recently been demonstrated.\cite{Stock2012}

If a semiconductor quantum dot is excited from its electronic ground state, the lowest excited configurations are the exciton states with one electron-hole pair in the system. Through further excitation from either of the excitons, the biexciton state with two electron-hole pairs can be excited (cf. Fig.~1). These excited states are relatively long-lived with lifetimes typically on the order of a nanosecond such that optical transitions can be studied in detail and also photon emission can be utilized efficiently. Most previous studies have focused on the emission of one or two subsequent (cascaded) photons from the biexciton to exciton or exciton to ground-state transition, respectively.\cite{Jayakumar2013,Mueller2014} In contrast to this cascaded emission, recently it was noted that semiconductor quantum dots can also efficiently couple to an optical light field through a direct two-photon transition from ground state to biexciton and vice versa.\cite{Brunner1994} Both of these states are spin-zero states such that a direct two-photon transition is allowed and efficient. Fully stimulated coherent two-photon excitation has been demonstrated in both degenerate\cite{Stufler2006} and two-color\cite{Boyle2010} scenarios. On the other extreme, a fully spontaneous two-photon emission was reported\cite{Ota2011} and explored.\cite{Valle2011b,Schumacher2012OptExpr}

\begin{figure}[b]
\hspace*{-3mm}\includegraphics[angle=0, scale=0.33, trim= 0mm 35mm 0mm 45mm, clip]{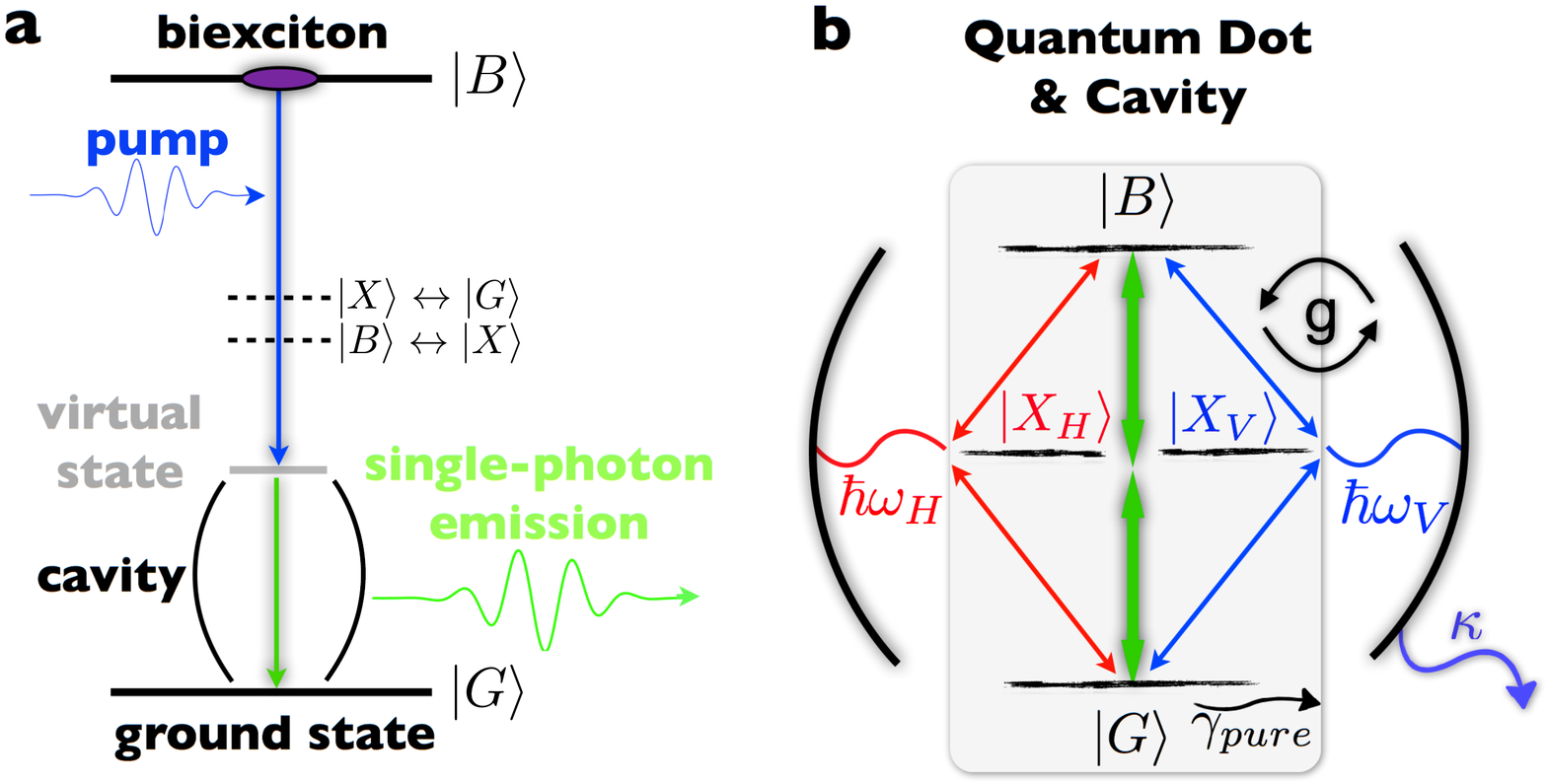}
\caption{\textbf{Single-photon generation through two-photon emission from a quantum-dot biexciton.} \textbf{a} The first photon is triggered by a laser field in a stimulated emission process into a virtual level inside the bandgap. The spontaneously emitted second photon is channeled into a cavity mode and has properties such as polarization and frequency complementary to the stimulating laser light field. \textbf{b} Illustration of the theoretical model for the quantum-dot cavity system analyzed. Details are given in the text.}
\end{figure}


\begin{figure*}
\hspace*{-2mm}\includegraphics[angle=0, scale=1.25]{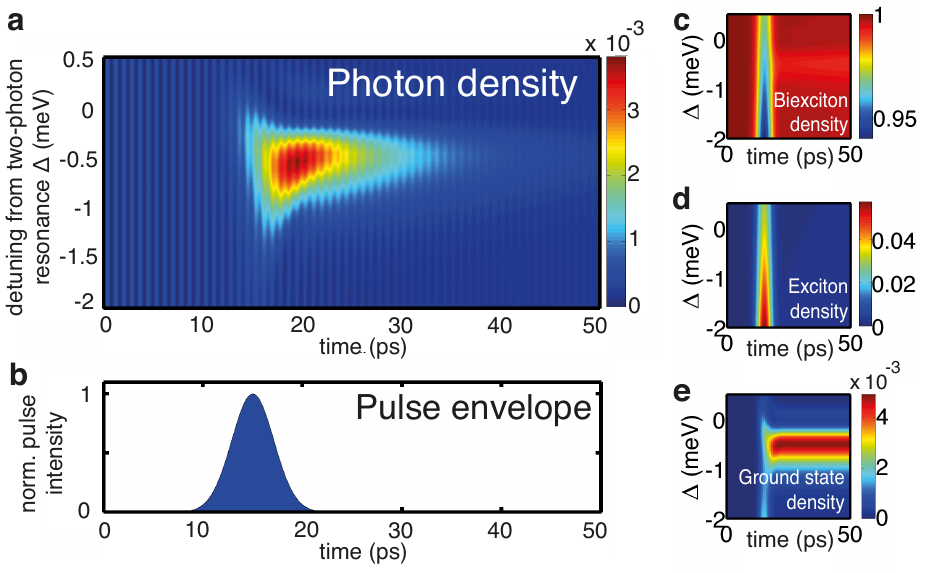}
\caption{\textbf{All-optical control of the single-photon emission event.} \textbf{a} Computed time-resolved photon density inside the cavity for different detunings of the laser pulse from the two-photon resonance condition. The temporal profile of the laser pulse is shown in \textbf{b} and biexciton, exciton, and ground state densities in \textbf{c}-\textbf{e}, respectively.}
\end{figure*}

In this Letter we analyze a mixed scenario for the two-photon emission from the biexciton: one photon is stimulated, the other one spontaneously emitted as illustrated in Fig.~1a. In analogy to a partially stimulated down-conversion process,\cite{Lamas2001} the first photon is triggered/stimulated by an external pump laser field. This field is off-resonant to all one-photon active transitions and drives the system between the biexciton state and a virtual level inside the system's band gap. However, the first photon only gets actually emitted if a second photon is spontaneously emitted (in our setup into a cavity mode) and bridges the remaining energy gap to the ground state. Given the first photon's stimulated nature, its properties are determined by the pump. Following from fundamental energy and spin conservation, the second photon has complementary properties such as polarization and frequency. Therefore changes in the parameters of the pump laser allow for all-optical control and on-the-fly changes to the properties of the emitted single photon.

Here we present a detailed theoretical analysis for a quantum-dot cavity system. We show numerically that a single-photon source as discussed above can be realized for a wide range of realistic system parameters. Our calculations show that the properties of the emitted photon (as a true quantum object) can indeed be all-optically controlled with the classical pump laser field.  Furthermore, we show that for excitation with a picosecond pump, the time of the emission event can be chosen to be inside the short time window marked by the presence of the pump pulse. 


The microscopic many-particle theory we use in our analysis is based on the quantum-dot cavity model illustrated in Fig.~1b. Included are the relevant electronic configurations of the quantum dot. These are ground state $|G\rangle$, excitons $|X_H\rangle$ and $|X_V\rangle$, and biexciton $|B\rangle$. The electronic system is coupled to the photons in two cavity modes with orthogonal polarizations and frequencies $\omega_{H,V}$. In addition to the quantized light fields in the cavity modes an off-resonant coherent laser field is included to trigger the photon emission. For a given initial state and given external laser field, we  evaluate the dynamical evolution of all occupations and coherences in the system. In particular, we extract detailed information about the photon emission from the system. Photon emission from the cavity and loss of electronic coherences on the relevant timescales are included. More details on the theoretical approach are given in the Methods Section below.

To study the scheme outlined above and illustrated in Fig.~1a, initially we prepare the quantum dot in the biexciton state with no photons in the cavity. Recent studies have shown the robust initialization of the biexciton.\cite{Mueller2014,Glaessl2013} Then a picosecond light pulse is applied, driving the system between the biexciton and a virtual state inside the band gap. When the pulse frequency $\omega_L$ is tuned such that the two-photon resonance condition from ground to biexciton state is fulfilled, $E_B-E_G=\hbar(\omega_L+\omega_{H,V})$, energy conservation allows spontaneous emission of a single photon into the cavity mode. To increase the probability of the emission event to occur during the presence of the pulse, first we use a high-quality cavity with $\kappa=\hbar/10\,\mathrm{ps^{-1}}$ (quality factor $Q\approx21000$ at $\hbar\omega=1.5\,\mathrm{eV}$) and a coupling strength of the single-photon transitions to the cavity mode of $g=\hbar/10\,\mathrm{ps^{-1}}\approx66\,\mathrm{\mu eV}$. For these parameters, Fig.~2a shows the computed time-resolved photon density inside the V-polarized cavity mode for different detuning of the V-polarized pump pulse from the two-photon resonance condition. The envelope of the $5\,\mathrm{ps}$ pulse with {peak Rabi energy of $1.5\,\mathrm{meV}$} is depicted in Fig.~2b. The result in Fig.~2a clearly shows that a photon is emitted when -- and only when -- the (here non-degenerate) two-photon resonance condition for laser pulse and cavity mode at $\Delta\approx0$ is met. Furthermore, photon emission only occurs during the presence of the laser pulse. This is evidenced by the dynamical built-up of a finite photon density inside the cavity mode when the pulse is switched on. After the pulse has passed, the remaining photons get emitted from the cavity on the $10\,\mathrm{ps}$ timescale determined by the cavity loss $\kappa$. We observe a slight field-induced shift of the emission in Fig.~1a from the ideal two-photon resonance condition at $\Delta=0$. In addition to the main emission feature, weak oscillations are visible as vertical stripes in Fig.~2a from the emission from the biexciton-exciton transition into the off-resonant cavity mode. 

The dynamics of the biexciton, exciton, and ground state densities are depicted in Fig.~2c-e. Clearly visible is the adiabatic following of these densities during the presence of the off-resonant pump pulse. However, only for the frequency window close to the two-photon resonance condition, the ground state and biexciton densities are changed by a sizeable amount after the pump pulse has passed. We stress that not a significant amount of density is generated in the exciton state(s) as the emission is strongly dominated by the direct two-photon channel. 

We note that the absolute probability for the photon emission to occur and with it the potential brightness of the source can be further optimized by increasing pump intensity and pulse length along with other system parameters such as biexciton binding energy, cavity frequency, coupling strength and cavity quality. We further note that for higher pumping intensities and stronger couplings the system dynamics can also become more complex. True benchmarking of the potential performance will be left for a future study possibly along with an experimental demonstration. We note that for the system parameters chosen here, by increasing the pulse length by a factor of $10$ to $50\,\mathrm{ps}$, we find an almost linear increase of the emitted photon density with pulse length while only generating an insignificant amount of exciton density.  

\begin{figure}
\includegraphics[angle=0, scale=1.5]{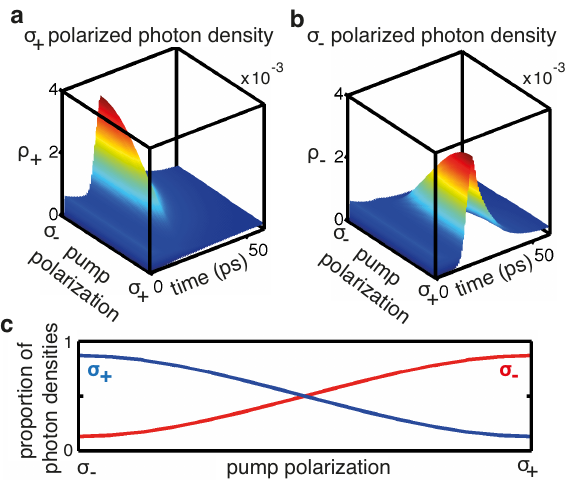}
\caption{\textbf{Optical polarization control of the single-photon emission.} Pump frequency is tuned to the {maximum emission at $\Delta=-0.54\,\mathrm{meV}$} in Fig.~2. Shown are the time-resolved \textbf{a} left- and \textbf{b} right-circularly polarized photon densities  $\rho_+$ and $\rho_-$ inside  the cavity for different polarization states of the pump pulse triggering the emission. \textbf{c} Time-averaged proportions of $\rho_+$ and $\rho_-$.}
\end{figure}

In Fig.~3 we demonstrate that with the polarization state of the pump field, the polarization state of the emitted photon can be controlled. The pump electric field amplitude and polarization state is parametrized as $\vec{E}=E_0\cdot(\cos{(p)}\vec{\sigma}_+ +\sin{(p)}\vec{\sigma}_-)$, here with the real-valued parameter $p$ with $p=0$ ($p=\pi/2$) corresponding to the limiting case of circularly polarized light $\sigma_+$ ($\sigma_-$). Parameters are the same as in Fig.~2. The two limiting cases are the circular polarization states $\sigma_+$ and $\sigma_-$: according to spin selection rules, when the pump is $+$ circularly polarized, the emitted photon is $-$ circularly polarized and vice versa. Figure~3c shows the time-averaged (averaged from $12$ to $45\,\mathrm{ps}$) proportions of photon densities in the circularly polarized states. With changing the pump polarization parameter $p$ the expected (almost sinusoidal) change in the polarization state is found. The achievable contrast is slightly reduced by the spontaneous biexciton decay through the exciton states. These results give unambiguous evidence that the emitted photon stems from a spontaneous emission into the cavity mode and can not result from photons pumped into the system through the laser source. 
By changing the polarization state of the pump anywhere in between the $+$ and $-$ circular polarization state (in general elliptically polarized), any polarization state can be realized for the emitted single photon of interest. Therefore, the classical laser field of the pump can be used to control the polarization state of a single photon as a true quantum object. 

In a scenario where also frequency filtering is applied during photon detection, the contrast for this polarization control could reach near $100\,\mathrm{\%}$ (no background photons are emitted in the spectral range of interest). To achieve efficient polarization control, the cavity modes must be degenerate (as can be realized in a micro pillar cavity for example), however, no frequency fine-tuning of cavity modes for example through temperature is needed as the stimulating laser can be tuned as needed.

In Fig.~4 we demonstrate that the scheme proposed here does not rely on the strong coupling or the high quality of the cavity mode used above. We present results obtained with a much lower coupling strength and in a low-quality cavity with {$g/\kappa=0.04$ with $g=1/50\,\mathrm{ps^{-1}}\approx13\,\mathrm{\mu eV}$} and $\kappa=1/2\,\mathrm{ps^{-1}}$ ($Q\approx4200$). As shown in Fig.~4, in this case the cavity resonance condition is alleviated such that also the two-photon resonance is smeared out. This leads to a significant emission of the photon for a wider range of pulse frequencies. We also find that with the photons not being stored in the low-quality cavity for a long time, the photon density (which in this case roughly correlates with the actual emission dynamics) occurs inside the cavity only during the presence of the short $5\,\mathrm{ps}$ pulse with {peak Rabi energy of $1.5\,\mathrm{meV}$}. This allows us to control the time of emission of the photon generated. This is explicitly shown in Fig.~4b and c for three different arrival times of the pump pulse. In contrast, using the usual cascaded emission to generate single photons, the time of the emission event can not be controlled but is random inside a typically much wider (nanosecond) time window. In Fig.~4b a background in photon density is observed from the competing gradual decay of the biexciton through the biexciton to exciton transition. If desired, using spectral filtering these background photons can be eliminated in the detection.

Whereas in the presence of the cavity mode, frequency control of the emitted photon (which is inherent to the process studied) can only be efficient inside the width of the cavity line, emission into free space within the same scheme would completely eliminate the cavity resonance condition. In this case the emission frequency could be all-optically tuned with the frequency of the pump. The available spectral range would be limited by other exciton resonances (higher exciton shells and LO-phonon assisted processes) present in the system. Even for emission into free space triggered by a sufficiently strong pump beam in a continuous wave experiment, the probability for the single photon being emitted from the desired two-photon process is expected to be on the same scale as the photon emission through the first order single photon transition from biexciton to exciton. Finally, we note that in this proof-of-principle study, we analyze emission into a cavity mode to increase the probability for the desired photon to be emitted during pulsed excitation. However, we would like to re-iterate that the general scheme does not rely on the presence of the cavity mode. Eventually, system design will be guided by the specific application in mind and its requirements. A choice will have to be made between using a high-quality cavity setup or a low-quality cavity or free-space emission. In the former case the probability for the emission of the desired photon during each pulse can be maximized, increasing brightness and potential on-demand character of the source. In the latter case the emission can be timed more accurately and the frequency and polarization state can be most flexibly tuned independent of any cavity resonance conditions. Also, a reasonable tradeoff needs to be made for the detuning of the cavity mode and pump from the single-photon active transitions. The further detuned, the more exclusively the two-photon transition can be utilized and studied. However, this is at the loss of the resonance enhancement of the two-photon transition when in the vicinity of the single-photon resonances.

\begin{figure}
\hspace*{-0mm}\includegraphics[angle=0, scale=1.5]{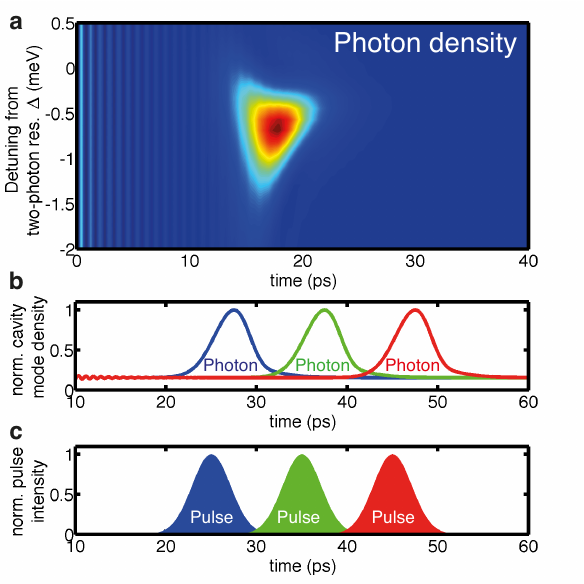}
\caption{\textbf{Controlling the time of emission with the arrival time of the pump.} A lower-quality cavity with lower coupling strength with $g/\kappa=0.04$ is used. \textbf{a} The $5\,\mathrm{ps}$ pump pulse is centred at $15\,\mathrm{ps}$. For a broad range of pump frequencies, a built-up of photon density and with it photon emission from the cavity is observed. In \textbf{b} we demonstrate that the timing of the photon emission can be controlled by varying the arrival time of the pump. For a pump detuning of {$\Delta=-0.54\,\mathrm{meV}$}, the normalized photon density is shown for pump arrival at $25\,\mathrm{ps}$, $35\,\mathrm{ps}$, and $45\,\mathrm{ps}$, respectively. \textbf{c} Pump envelope corresponding to panel \textbf{b}.}
\end{figure}

In conclusion, we have proposed and theoretically analyzed a new scheme for single-photon generation with semiconductor quantum dots. This scheme utilizes a partially stimulated two-photon emission from the quantum-dot biexciton. In this scheme, properties (polarization state, frequency, time of emission) of the spontaneously emitted single photon can be controlled on-the-fly and all-optically with a classical laser pulse. Future possibilities include using the same scheme for emission into free space such that frequency can be controlled in a wider spectral range. Applying a chirp in frequency and/or polarization state of the pump to control the complex temporal dynamics of the single-photon emission event could also be of great interest. An experimental realization of the proposed scheme would be highly desirable. We envision that the scheme we propose bears great promise for realization of the next generation of versatile quantum-dot based single-photon sources.

\section{Methods}

We include the relevant electronic configurations of the quantum dot in our theory. These are ground state $|G\rangle$, excitons $|X_H\rangle$ and $|X_V\rangle$, and biexciton $|B\rangle$. The electronic system is then coupled to the photons in two cavity modes with orthogonal polarizations and frequencies $\omega_{H,V}$. In addition to the quantized light fields in the cavity modes an off-resonant coherent laser field is included to trigger the emission. In rotating wave approximation, the many-particle system Hamiltonian then reads:
\newcommand{\pro}[1]{\ensuremath{|#1\rangle\langle #1|}}
\newcommand{\proj}[2]{\ensuremath{|#1\rangle\langle #2|}}
\begin{align}
\mathcal{H} &=E_G \pro{G}+E_B\pro{B} + \sum_{i=H,V}{E_i \pro{X_i}} \nonumber  \\  & + \sum_{i=H,V}{\Big(\hbar\omega_ib_i^\dag b_i-\big[g\big(\proj{G}{X_i} b_i^\dag +\proj{X_i}{B}b_i^\dag\big) + \mathrm{h.c.}\big]\Big)} \nonumber \\ & + \sum_{i=H,V}{\Big(\big(\proj{G}{X_i}\Omega_i^\ast(t) +\proj{X_i}{B}\Omega_i^\ast(t)\big) + \mathrm{h.c.}\Big)} \nonumber \,.
\end{align}
Here, $b^\dag_i$ ($b_i$) denote creation (annihilation) operators of photons in the cavity modes $H$ and $V$ and $\Omega_i(t)$ gives the Rabi energy of the time dependent coherent laser field projected onto the respective transitions with $i$-polarization. As for the validity of our hybrid theory including both classical and quantized light fields, we note that it is important that in all our evaluations below, the laser field is off-resonant to the cavity modes by several $\mathrm{meV}$. The coupling strength of the electronic system to cavity modes is given by $g$. The time-evolution of the system density operator $\rho_s$ obeys the following equation of motion:
\begin{align}\label{Eq_motion}
\frac{\partial}{\partial t}\rho_s=-\frac{i}{\hbar}[\mathcal{H},\rho_s]+\mathcal{L}_{\text{cavity}}(\rho_s)+\mathcal{L}_{\text{pure}}(\rho_s)\,.
\end{align} 
Coupling of the system to the environment is included through the two dissipative terms $\mathcal{L}_{\text{cavity}}(\rho_s)$ and $\mathcal{L}_{\text{pure}}(\rho_s)$. The finite lifetime $\hbar/\kappa$ of the photons inside the cavity is taken into account through
the Lindblad term
\begin{align} 
\mathcal{L}_{\text{cavity}}(\rho_s)=\frac{\kappa}{2}\sum_{i=H,V}(2b_i\rho_s
b_i^\dag-b_i^\dag b_i \rho_s- \rho_s b_i^\dag b_i)\,.
\end{align} 
A pure dephasing of coherences between electronic configurations is included through
\begin{align}
\mathcal{L}_{\text{pure}}(\rho_s)=-\frac{1}{2}\sum_{\chi,\chi',\chi\neq\chi'}{\gamma_{\chi\chi'}^{\text{pure}}}{}\pro{\chi}\rho_s\pro{\chi'}\,,
\end{align}
with $\chi,\chi'\in\{G,X_H,X_V,B\}$.\cite{Troiani2006} We use the same value
{$\gamma^{\text{pure}}_{\chi\chi'}=\gamma=\hbar/200\,\mathrm{ps^{-1}}\approx 3\,\mu$eV} for all electronic coherences, which is a realistic value for pure dephasing of excitonic coherences at low
temperature.\cite{Laucht2009} We note that our results are qualitatively robust even with a pure dephasing much faster than assumed here. Fine structure splitting between exciton levels -- typically of the order of several tens of $\mathrm{\mu eV}$  -- does only cause minor quantitative changes to the results and with $E_H=E_V$ is set to zero for simplicity. Degeneracy is assumed for the cavity modes, $\omega_H=\omega_V$, which is needed for efficient polarization control. A reasonable biexciton binding energy of $3\,\mathrm{meV}$ is assumed and the cavity modes are tuned $5\,\mathrm{meV}$ to the red of the biexciton to exciton transitions. In this work we assume the system to be initially in the biexciton configuration with no photons inside the cavity. For this initial condition, the system dynamics is obtained by explicitly solving Eq.~(\ref{Eq_motion}) in the finite-dimensional Fock-space spanned by the degrees of freedom of our system. Expectation values of any operator $\hat{A}$ are computed by taking the trace with the system density operator, $\langle\hat{A}\rangle=\mathrm{tr}\{\hat{A}\rho_s\}$.  

We note that for simplicity some secondary effects that could slightly change the results have not been considered: phonon-assisted transitions into cavity mode or laser field are expected to be weak for the detuning of several $\mathrm{meV}$ used;\cite{Roy2011} the radiative loss into other off-resonant cavity modes is expected to play a small quantitative role on the timescales studied.
 
\section{Acknowledgments}

We gratefully acknowledge financial support from the DFG through SFB TRR142, from the BMBF through Q.com 16KIS0114, and a grant for computing time at $\mathrm{PC^2}$ Paderborn Center for Parallel Computing. We acknowledge valuable discussions on the subject matter with Rolf Binder and Dominik Breddermann.

\end{document}